\documentclass[conference,10pt]{IEEEtran}

\IEEEoverridecommandlockouts

\usepackage{amsmath,amsthm,relsize}
\usepackage{amsfonts, amssymb, cuted}
\usepackage{cleveref}
\usepackage{mathtools}
\usepackage{algorithm}
\usepackage[noend]{algpseudocode}
\usepackage{cite}
\usepackage{xcolor}
\usepackage{soul}
\usepackage{graphicx}
\usepackage{pgfplotstable}
\usepackage{pgfplots}
\usepackage{nicefrac}
\usepackage{enumitem}
\usepackage{support-caption}
\usepackage[style=base,font=footnotesize,skip=1pt,belowskip=-4pt]{caption}
\usepackage{multirow}
\pgfplotsset{compat=1.14}

\theoremstyle{definition}

\newcommand{\CA}[0]{{\mathcal{A}}}
\newcommand{\CB}[0]{{\mathcal{B}}}

\newcommand{\CF}[0]{{\mathcal{F}}}

\newcommand{\CN}[0]{{\mathcal{N}}}

\newcommand{\CS}[0]{{\mathcal{S}}}
\newcommand{\CT}[0]{{\mathcal{T}}}

\newcommand{\SfA}[0]{{\mathsf{A}}}

\begin{document}

\title{Coded Caching with Uneven Channels:\\A Quality of Experience Approach
\thanks{This work was supported by the Academy of Finland under grants no. 319059 (Coded Collaborative Caching for Wireless Energy Efficiency) and 318927 (6Genesis Flagship).}}
\date{January 2020}

\author{%
  \IEEEauthorblockN{MohammadJavad Salehi and Antti T\"olli}
  \IEEEauthorblockA{Center for Wireless Communications (CWC)\\
                    University of Oulu\\ 
                    90570 Oulu, Finland\\
                    Email: \{first\_name.last\_name\}@oulu.fi}
  \and
  \IEEEauthorblockN{Seyed Pooya Shariatpanahi}
  \IEEEauthorblockA{School of Electrical and Computer Engineering\\
                    College of Engineering, University of Tehran\\
                    Tehran, Iran\\
                    Email: p.shariatpanahi@ut.ac.ir}

}

\maketitle

\begin{abstract}
The rate performance of wireless coded caching schemes is typically limited
by the lowest achievable per-user rate in the given multicast group, during each transmission time slot. In this paper, we provide a new coded caching scheme, alleviating this worst-user effect for the prominent case of multimedia applications. In our scheme, instead of maximizing the symmetric rate among all served users, we maximize the total quality of experience (QoE); where QoE at each user is defined as the video quality perceived by that user. We show the new scheme requires solving an NP-hard optimization problem. Thus, we provide two heuristic algorithms to solve it in an approximate manner; and numerically demonstrate the near-optimality of the proposed approximations. Our approach allows flexible allocation of distinct video quality for each user, making wireless coded caching schemes more suitable for real-world implementations.
\end{abstract}


\section{Introduction}
Network data volume has continuously grown during the past years. The global IP (Internet Protocol) data volume is expected to exceed 4.8 Zettabytes ($10^{21}$ bytes) by 2022, from which 71 percent will pass through wireless networks~\cite{cisco2018cisco}. Mobile video applications account for a major part of this data volume; as their share is expected to reach 79\% of the global mobile data traffic by 2022. Consequently, great efforts are made by the research community, for developing new communication schemes well-suited to current and future (e.g. immersive viewing) video applications.

Most video applications share a few important features. First, the source of the received content is not important, as long as each user receives its requested content \cite{jacobson2009networking}. Moreover, the content request probability distribution is uneven and there is a prime time where the request rate is higher \cite{cha2009analyzing}. As a result, caching popular content is considered as a viable solution for large scale video delivery \cite{jacobson2009networking,salehi2017optimality}; specially considering the declining price of memory chips \cite{leventhal2008flash}.

Recently, Coded Caching (CC) is introduced as a promising extension to conventional caching schemes \cite{maddah2014fundamental}. It enables a global caching gain, proportional to the total cache size in the network, to be achieved in addition to the local caching gain at each user. This extra gain is enabled by multicasting carefully designed codewords to various user subsets, such that each codeword contains useful data for all users in its target subset. It is also shown that CC gain is additive with the spatial gain of using multiple antennas \cite{shariatpanahi2016multi,shariatpanahi2018physical}; making CC even more desirable for future networks in which multi-input, multi-output (MIMO) communications play a major role \cite{6Genesis2019KeyIntelligence}.

Following~\cite{maddah2014fundamental}, a significant effort has been carried out by the research community to make CC implementation practical for future networks.
For example, subpacketization, defined as the number of fragments each file should be split into for a CC scheme to work properly, is thoroughly investigated in \cite{lampiris2018adding,salehi2019subpacketization,salehi2019coded}; while CC performance at low-SNR wireless communications is studied in \cite{tolli2017multi,salehi2019beamformer}. Unfortunately, the multicast nature of CC makes its performance to be compromised if served users have diverse channel conditions. In fact, if the channel capacity is small for a specific user $k$, the achievable multicast rate of any user subset including $k$ will be limited to the rate of $k$. This issue, known as the worst-user effect, is addressed in \cite{lampiris2019fundamental,destounis2017alpha}. In \cite{lampiris2019fundamental} a superposition coding approach is used, in which more transmit power is allocated to the weaker user. On the other hand, a dynamic network is considered in \cite{destounis2017alpha} where queuing techniques are used to transmit more data to the stronger users during the time.

In this paper, we take a new approach to the worst-user effect by optimizing the total Quality of Experience (QoE) of all users during a transmission interval (defined by application requirements). Considering a single-antenna communication setup, we use Multiple Descriptor Codes (MDC), introduced in \cite{goyal2001multiple}, to enable flexible video quality at each user. Expressing QoE as the number of MDC elements a user receives (i.e. the video quality it experiences), we propose a partial CC scheme for optimizing the QoE sum. We show the optimization problem is NP-hard; and provide two heuristic algorithms to solve it approximately.
Our scheme provides a new viewpoint for solving the worst-user effect in CC schemes, enabling CC to be better tailored for future wireless networks.

Throughout the text, we use $[K]$ to denote  $\{1,2,...,K\}$ and $[i:j]$ to represent $\{i,i+1,...,j\}$. Sets are denoted by calligraphic letters. For two sets $\CA$ and $\CB$, $\CA \backslash \CB$ is the set of elements in $\CA$ which are not in $\CB$; and $|\CA|$ represents the number of elements in $\CA$.

\section{System Model}
\subsection{Coded Caching (CC)}

We consider a CC setup similar to \cite{maddah2014fundamental}, where a single server communicates with $K$ users over a shared broadcast link. Each user is equipped with a cache memory of size $Mf$ bits; and requests files from a library $\CF$, where $|\CF| = N$ and each file $W \in \CF$ has the same size of $f$ bits. For simplicity, we assume a normalized data unit and drop $f$ in subsequent notations.

The system operation consists of two phases. During the placement phase, which takes place at low network traffic time and without any knowledge of file request probabilities in the future, each user stores data chunks of files in $\CF$, in its cache memory. Following \cite{maddah2014fundamental}, we define the CC gain as $t = \frac{KM}{N}$ and assume $t$ is an integer. Moreover, we assume each file $W \in \CF$ is split into $P = \binom{K}{t}$ equal-sized chunks $W_{\CT}$, where $\CT$ can be any subset of users with $|\CT| = t$. Then during the placement phase, each user $k \in [K]$ stores data chunks $W_{\CT}$, for every $W \in \CF$ and $\CT \ni k$, in its cache memory.

At the beginning of the delivery phase, every user $k \in [K]$ reveals its requested file $W(k) \in \CF$.
Based on users' requests and in accordance with a delivery algorithm, the server builds carefully designed codewords; and transmits each codeword in a separate time slot over the shared communication channel. Each user $k \in [K]$ should be able to decode $W(k)$, using data stored in its cache memory together with the data received from the channel.
For the original scheme of \cite{maddah2014fundamental}, a codeword $X(\CS)$ is built for every $\CS \subseteq [K]$ with $|\CS| = t+1$. Denoting the bit-wise XOR operation with $\oplus$, $X(\CS)$ is built as
\begin{equation}
\label{eq:codeword_mn}
    X(\CS) = \bigoplus\limits_{k \in \CS} W_{\CS \backslash \{k\}}(k) \; .
\end{equation}
In \cite{maddah2014fundamental} it is shown that
after the transmission of $X(\CS)$ is concluded, every user $k \in \CS$ can remove unwanted terms using its cache contents; and decode $W_{\CS \backslash \{k\}}(k)$ interference-free.
This will decrease the load on the shared link by a factor of $t+1$, compared to an uncoded, unicast transmission strategy.

\subsection{The Worst-User Effect}
In \cite{maddah2014fundamental} it is assumed that the channel capacity for all users is one (normalized) data unit per channel use. For a more realistic setup, we assume after $X(\CS)$ is transmitted, user $k$ receives $y_k(\CS) = h_k^T X(\CS) + z_k$, where $h_k \in \mathbb{C}$ and $z_k \sim \CN(0,N_0)$ denote the channel coefficient and the additive Gaussian noise at user $k$, respectively. Based on this assumption, for user $k \in \CS$ to be able to decode $X(\CS)$, the transmission rate of $X(\CS)$ should be smaller than or equal to the channel capacity of user $k$; which is denoted by $c_k$ and calculated as
\begin{equation}
\label{eq:user_capacity}
    c_k = \log (1 + \frac{P_T |h_k|^2}{N_0}) \; ,
\end{equation}
where $P_T$ stands for the available transmission power. However, for the delivery algorithm to work properly, every user $k \in \CS$ should be able to decode $X(\CS)$; which means for the transmission rate of $X(\CS)$, denoted by $r(\CS)$, we should have
\begin{equation}
\label{eq:codeword_rate_mn}
    r(\CS) \le \min_{k \in \CS} c_k = \min_{k \in \CS} \log (1 + \frac{P_T |h_k|^2}{N_0}) \; .
\end{equation}
Assuming the transmission is carried out with the highest rate and considering the fact each data part has the size of $\frac{1}{P}$ data units, the delivery time for $X(\CS)$ is then calculated as
\begin{equation}
    T(\CS) = \frac{1}{P} \frac{1}{\min_{k \in \CS} c_k} \; ,
\end{equation}
and the total delivery time of all users would be equal to
\begin{equation}
\label{eq:total_time_mn}
    T_T = \sum_{\CS} T(\CS) = \frac{1}{P} \sum_{\CS} \frac{1}{\min_{k \in \CS} \log (1 + \frac{P_T |h_k|^2}{N_0})} \; .
\end{equation}

Equation \eqref{eq:total_time_mn} indicates that $T_T$ becomes very large, in case $|h_k|$ is very small for some user $k \in [K]$. Although the achievable $T_T$ is still smaller compared with an uncoded strategy, the large delivery time can be undesirable for users with better channel conditions (users with larger $|h_k|$ values); as these users would have experienced smaller delivery times if they had received their requested data through an uncoded, unicast transmission. In fact, coded caching causes the download rate of all users to be limited to the worst achievable rate, known as the \textit{worst-user effect} in the literature \cite{lampiris2019fundamental}. Clearly, this issue is more prominent if the ratio between the largest and smallest values of $|h_k|$ is larger.

\subsection{A Quality of Experience (QoE) Approach}

In order to address the worst-user issue, we introduce a new approach for designing coded caching schemes; which is based on the QoE definition and is well-suited for the prominent case of video-based applications. The core of this new design approach is based on using Multiple Descriptor Codes (MDC), as introduced in \cite{goyal2001multiple}. %
MDC enables creating multiple descriptors of the same video (or any other multimedia) file, such that any single descriptor is enough for reconstructing a basic-quality replica of the original file; and the quality is increased as more descriptors are used during the reconstruction process.

Let us assume file library $\CF$ includes video files only; and instead of denoting a file fragment,
each $W_{\CT}$ represents one of the $P = \binom{K}{t}$ descriptors of the file $W \in \CF$, with size $\frac{1}{P}$ data units. Similar to the original CC scheme, $\CT$ can be any subset of users with $|\CT| = t$; and at the cache memory of user $k \in [K]$ we store $W_{\CT}$, for every $W \in \CF$ and $\CT \ni k$. Using this scheme, each user $k \in [K]$ is able to decode a basic quality of its requested video file $W(k)$ using its cache contents; and the quality increases as it gets more descriptors from the server. Defining QoE at user $k$ as the total number of $W(k)$ descriptors available at user $k$ after the transmission is completed, the question is how much the QoE sum at all users can be improved, during a limited transmission time.


In order to formulate this problem, we first need to revise the delivery algorithm. Similar to the original CC scheme, we select all subsets $\CS \subseteq [K]$ with $|\CS| = t+1$; and label the users in set $\CS$ as $k(\CS,1)$, $k(\CS,2)$, ..., $k(\CS,t+1)$, such that if $1 \le i < j \le t+1$, $|h_{k(\CS,i)}| \ge |h_{k(\CS,j)}|$. Then instead of building $X(\CS)$ as \eqref{eq:codeword_mn}, we build the codeword $Y_{j_{\CS}}(\CS)$, $j_{\CS} \in [0:t+1]$, using
\begin{equation}
    Y_{j_{\CS}}(\CS) = \bigoplus\limits_{i \in [j_{\CS}]} W_{\CS \backslash \{k(\CS,i)\}}\big( k(\CS,i) \big) \; .
\end{equation}
In other words, we take the first $j_{\CS}$ users of $\CS$ with better channel conditions; and create the codeword using the descriptors requested by these users only. Clearly, the maximum error-free transmission rate for $Y_{j_{\CS}}(\CS)$ is equal to
\begin{equation}
    c(\CS,j_{\CS}) = c_{k(\CS,j_{\CS})} = \log \Bigg(1 + \frac{P_T \big|h_{k(\CS,j_{\CS})}\big|^2}{N_0}\Bigg) \; ,
\end{equation}
where, compared with \eqref{eq:codeword_rate_mn}, the minimizing operation is removed as the users are sorted and hence $k(\CS,j_{\CS})$ has the worst channel condition among the target users. In fact, transmitting $Y_{j_{\CS}}(\CS)$ instead of $X(\CS)$ enables $j_{\CS}$ descriptors to be delivered with rate $c(\CS,j_{\CS})$; instead of $t+1$ descriptors with rate $c(\CS)$ (as in the original CC scheme). The question is then how to select $j_{\CS}$ values, such that QoE sum is maximized.

Let us denote the transmission time limit by $T_{\lim}$. Assuming for every $\CS$, $Y_{j_{\CS}}(\CS)$ is transmitted with the highest possible rate $c(\CS,j_{\CS})$, the delivery time for $Y_{j_{\CS}}(\CS)$ would be
\begin{equation}
    T(\CS,j_{\CS}) = \frac{1}{P} \frac{1}{c(\CS,j_{\CS})} \; ,
\end{equation}
and the QoE sum optimization problem can be written as the integer-programming problem
\begin{equation}
\label{eq:optimization_problem_main}
\begin{aligned}
    &\max_{j_{\CS}} \; \sum_{\CS} j_{\CS} \; , \\
    &s.t. \quad \sum_{\CS} T(\CS,j_{\CS}) \le T_{\lim} \; .
\end{aligned}
\end{equation}

\subsection{Demonstrative Example}
Consider a small network of $K=5$ users with diverse channel conditions, $t=2$ and $P=\binom{5}{2} = 10$. Assume for every user $k \in [5]$ we have $c_k = \frac{1}{10}\frac{1}{k}$; i.e. delivering a video descriptor to user $k$ requires $k$ seconds. 
It can be verified that in this setup, each user has 4 descriptors of its requested file in the cache memory; and needs 6 other ones for the highest possible QoE. Moreover, uncoded delivery requires 90 seconds for all users to reach the highest QoE; while coded strategy of \cite{maddah2014fundamental} cuts this time half to 45 seconds. So in case the higher-layer application requires $T_{\lim} < 45$, it is not possible to serve every user with the highest QoE; and one needs to build the codewords such that the QoE sum is maximized for all users.

Let us consider the case $T_{\lim} = 10$ seconds. Solving the optimization problem \eqref{eq:optimization_problem_main} results in $j_{\CS}$ values
\begin{equation*}
    \begin{gathered}
        \CS_1 = \{1,2,3\} \rightarrow j_{\CS_1} = 3 \; , \; \CS_2 = \{1,2,4\} \rightarrow j_{\CS_2} = 2 \; , \\
        \CS_3 = \{1,2,5\} \rightarrow j_{\CS_3} = 2 \; , \; \CS_4 = \{1,3,4\} \rightarrow j_{\CS_4} = 1 \; , \\
        \CS_5 = \{1,3,5\} \rightarrow j_{\CS_5} = 1 \; , \; \CS_6 = \{1,4,5\} \rightarrow j_{\CS_6} = 1 \; , \\
        \CS_7 = \{2,3,4\} \rightarrow j_{\CS_7} = 0 \; , \; \CS_8 = \{2,3,5\} \rightarrow j_{\CS_8} = 0 \; , \\
        \CS_9 = \{2,4,5\} \rightarrow j_{\CS_9} = 0 \; , \; \CS_{10} = \{3,4,5\} \rightarrow j_{\CS_{10}} = 0 \; , \\
    \end{gathered}
\end{equation*}
which means, for example, we transmit $W_{\{2,5\}}(1) \oplus W_{\{1,5\}}(2)$ with rate $c_2$ for users in $\CS_3$; while for users in $\CS_6$ we transmit $W_{\{4,5\}}(1)$ with rate $c_1$. Clearly, using these codewords, the total QoE of users in this network becomes 10 (note that the current descriptors in the cache memories are not counted).

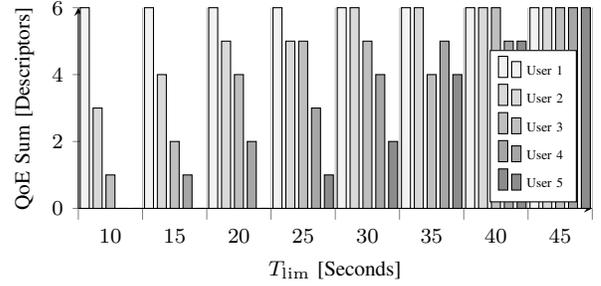
\begin{figure}[t]
    \centering
    {%
   
    \begin{tikzpicture}
    \begin{axis}
    [
    width = 0.95\columnwidth,
    height = 0.48\columnwidth,
    axis lines = left,
    xlabel = \smaller {$T_{\lim}$ [Seconds]},
    ylabel = \smaller {QoE Sum [Descriptors]},
    ylabel near ticks,
    legend pos = south east,
    legend style = {cells={align=center}},
    ticklabel style={font=\smaller},
    ybar interval=0.7,
    ]
    
    \addplot
    [black,fill = gray!10,]
    table
    [x=Tlim,y=U1]
    {Data/QoE_OverTime.txt};
    \addlegendentry{\tiny User 1}

    \addplot 
    [black,fill = gray!30,]
    table
    [x=Tlim,y=U2]
    {Data/QoE_OverTime.txt};
    \addlegendentry{\tiny User 2}
    
    \addplot 
    [black,fill = gray!50,]
    table
    [x=Tlim,y=U3]
    {Data/QoE_OverTime.txt};
    \addlegendentry{\tiny User 3}
    
    \addplot 
    [black,fill = gray!70,]
    table
    [x=Tlim,y=U4]
    {Data/QoE_OverTime.txt};
    \addlegendentry{\tiny User 4}
    
    \addplot[black,fill = gray!90,]
    table
    [x=Tlim,y=U5]
    {Data/QoE_OverTime.txt};
    \addlegendentry{\tiny User 5}
    
    \end{axis}
    \end{tikzpicture}
    }
    \caption{User-specific QoE versus $T_{\lim}$ for the example network}
    \label{fig:Tlim_effect_bar}
\end{figure}

\begin{figure}[t]
    \centering
    {%

    \begin{tikzpicture}

    \begin{axis}
    [
    width = 0.95\columnwidth,
    height = 0.48\columnwidth,
    axis lines = left,
    xlabel = \smaller {$T_{\lim}$ [Seconds]},
    ylabel = \smaller {QoE Sum [Descriptors]},
    ylabel near ticks,
    ticklabel style={font=\smaller},
    grid=both,
    major grid style={line width=.2pt,draw=gray!30},
    ]
    
    \addplot
    [black]
    table[y=QoE,x=Time]
    {Data/QoE_sum.txt};
    
    \end{axis}

    \end{tikzpicture}
    }

    \caption{QoE sum versus $T_{\lim}$ for the example network}
    \label{fig:Tlim_effect}
\end{figure}
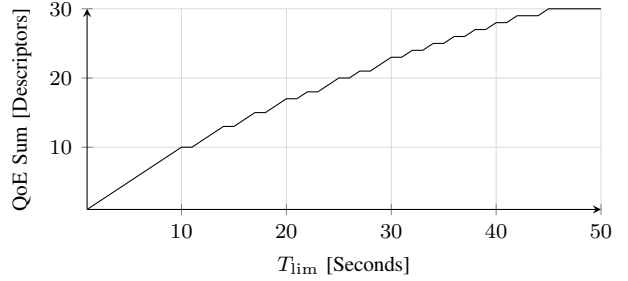

In Figure \ref{fig:Tlim_effect_bar} and \ref{fig:Tlim_effect} we have plotted user-specific and total QoE for this network, for various $T_{\lim}$ values. Clearly, both user-specific and total QoE increase with $T_{\lim}$. Specifically, at $T_{\lim} = 45$ seconds, total QoE reaches its largest value of 30, on par with the CC scheme of \cite{maddah2014fundamental}. Moreover, users with better channel conditions are prioritized (and enjoy higher QoE) at smaller $T_{\lim}$.


\section{QoE Maximization}
\subsection{Exhaustive Search}

The QoE sum optimization problem in \eqref{eq:optimization_problem_main} is a special case of the famous multiple choice knapsack problem \cite{sinha1979multiple}, in which every set $\CS$ represents a class from which we can select at most one of the $t+1$ items (indexed by $j_{\CS}$). Each item has a known weight, denoted by $T(\CS,j_{\CS})$; and the total weight of the selected items should be smaller than the knapsack capacity ($T_{\lim}$). This problem is known to be NP-hard to solve.

Using exhaustive search, one can find the optimal solution. A recursive procedure for the search operation is provided in Algorithm \ref{alg:exh}; in which $\Bar{\CS}$ denotes the set of all available $\CS$ sets and $Q(\Bar{\CS},T_{\lim})$ represents the optimal solution.
%
This procedure requires comparison of all $(t+2)^{\gamma}$ possible selections, where $\gamma = \binom{K}{t+1}$. This means computation complexity grows exponentially with respect to both $K$ and $t$, making the problem computationally intractable for even moderate $K$ and $t$. 

\setlength{\textfloatsep}{4pt}
\begin{algorithm}[ht]
\caption{Exhaustive Search Procedure}
\label{alg:exh}
\begin{algorithmic}[1]
    \Function{Exhaustive}{$\Bar{\CS}, T_{\lim}$}
        \State $Q \gets 0$
        \State Randomly Select $\CS_l \in \Bar{\CS}$
        \If{$|\Bar{\CS}| = 1$}
            \State $\hat{i} \gets 0$
            \ForAll{$i \in [t+1]$}
                \If{$T(\CS_l,i) \le T_{\lim}$ and $i > \hat{i}$}
                    \State $\hat{i} \gets i$
                \EndIf
            \EndFor
            \State $Q \gets \hat{i}$
        \Else
            \State $\gamma_0 \gets$ \textsc{Exhaustive}$\big(\Bar{\CS} \backslash \{\CS_l\}$, $T_{\lim}\big)$
            \ForAll{$i \in [t+1]$}
                \State $\gamma_i \gets$ \textsc{Exhaustive}$\big(\Bar{\CS} \backslash \{\CS_l\}$, $T_{\lim}-T(\CS_l,i)\big)$
                \State $\gamma_i \gets \gamma_i + i$
            \EndFor
            \State $Q \gets \max_{i \in [0:t+1]} \gamma_i$
        \EndIf
        \Return $Q$
    \EndFunction
\end{algorithmic}
\end{algorithm}

\subsection{SDT Approximation}

As a first approximation, we provide SDT (Step Delivery Time), as presented in Algorithm \ref{alg:std}. As a brief explanation, during each iteration we find the minimum increase in the total required transmission time, per one new descriptor being delivered. 
Auxiliary variable $\alpha(\CS)$ denotes the current $j_{\CS}$ value; and $\beta(\CS)$ indicates the increase in the required transmission time, if $\alpha(\CS)$ is increased by one. Complexity-wise, Algorithm~\ref{alg:std} requires at most $(t+1)\gamma$ iterations
; and at each iteration, a minimum is taken over a set of at most $\gamma$ numbers.

\begin{algorithm}[ht]
\caption{SDT Approximation Procedure}
\label{alg:std}
\begin{algorithmic}[1]
    \Function{RunSDT}{$\Bar{\CS}, T_{\lim}$}
        \State $Q \gets 0$
        \State $\hat{T} \gets T_{\lim}$
        \ForAll{$\CS \in \Bar{\CS}$}
            \State $\alpha(\CS) \gets 0$
            \State $\beta(\CS) \gets T(\CS,\alpha(\CS))$ 
        \EndFor
        \State $\hat{\CS} \gets \arg\min_{\CS} \beta(\CS)$
        \While{$\beta(\hat{\CS}) \le \hat{T}$}
            \State $Q \gets Q+1$
            \State $\hat{T} \gets \hat{T} - \beta(\hat{\CS})$
            \State $\alpha(\hat{\CS}) \gets \alpha(\hat{\CS}) + 1$
            \If{$\alpha(\hat{\CS}) = t+1$}
                \State $\beta(\hat{\CS}) \gets +\infty$
            \Else
                \State $\beta(\hat{\CS}) \gets T(\hat{\CS},\alpha(\hat{\CS})) - \beta(\hat{\CS})$
            \EndIf
             \State $\hat{\CS} \gets \arg\min_{\CS} \beta(\CS)$
        \EndWhile
        \Return $Q$
    \EndFunction
\end{algorithmic}
\end{algorithm}



\subsection{PDT Approximation}
Instead of finding the minimum increase in the required transmission time, PDT approximation, presented in Algorithm \ref{alg:pdt}, is based on finding the minimum Perceived Delivery Time (PDT) at each iteration. PDT is defined as the increase in the required transmission time, normalized by the number of new descriptors being delivered to the users in some set $\CS$. Formally, if instead of $j_{\CS}$ descriptors, we deliver $j_{\CS}'$ descriptors to the users of set $\CS$, PDT for this action is calculated as
\begin{equation*}
    \frac{T(\CS,j_{\CS}')-T(\CS,j_{\CS})}{j_{\CS}'-j_{\CS}} \; .
\end{equation*}
Let us use $\SfA(\CS,j_{\CS}')$ to denote the action of increasing the number of descriptors being delivered to the users in set $\CS$, from $j_{\CS}$ to $j_{\CS}'$. $\SfA(\CS,j_{\CS}')$ is \textit{feasible}, if it does not violate the total transmission time constraint ($T_{\lim}$). In PDT approximation, at each iteration, we find $\CS$ and $j_{\CS}'$ such that $\SfA(\CS,j_{\CS}')$ is feasible and has the minimum PDT among all feasible actions. 

It should be noted that PDT is more complex than SDT, for two good reasons. First, the search operation for finding the minimum increment in required transmission time in PDT is performed over a larger set of numbers, with approximately $t+1$ times more elements than SDT. Second, updating auxiliary variables is more complex in PDT; as one needs to calculate the perceived delivery time (which itself requires one subtraction and one division) for a large number of elements (compared with only one subtract operation in SDT).

\begin{algorithm}[t]
\caption{PDT Approximation Procedure}
\label{alg:pdt}
\begin{algorithmic}[1]
    \Function{RunPDT}{$\Bar{\CS}, T_{\lim}$}
        \State $Q \gets 0$
        \State $\hat{T} \gets T_{\lim}$
        \ForAll{$\CS \in \Bar{\CS}$}
            \State $\alpha(\CS) \gets 0$
            \ForAll{$i \in [t+1]$}
                \If{$T_{k(\CS,i)} \le \hat{T}$}
                    \State $\beta(\CS,i) \gets \nicefrac{T(\CS,i)}{i}$
                \Else
                    \State $\beta(\CS,i) \gets +\infty$
                \EndIf
            \EndFor
        \EndFor
        \State $(\hat{\CS},\hat{i}) \gets \arg\min_{\CS,i} \beta(\CS,i)$
        \While{$\beta(\hat{\CS}) \le \hat{T}$}
            \State $Q \gets Q+\hat{i}$
            \State $\hat{T} \gets \hat{T} - \beta(\hat{\CS},\hat{i}) \times (\hat{i}-\alpha(\hat{\CS})) $ 
            \State $\alpha(\hat{\CS}) \gets \alpha(\hat{\CS}) + \hat{i}$
            \ForAll{$\CS \in \Bar{\CS}$}
                \ForAll{$i \in [\alpha(\CS)]$}
                    \State $\beta(\CS,i) \gets +\infty$
                \EndFor
                \ForAll{$i \in [\alpha(\CS)+1:t+1]$}
                    \If{$T(\CS,i) - T(\CS,\alpha(\CS)) \le \hat{T}$}
                        \State $\beta(\CS,i) \gets \nicefrac{(T(\CS,i) - T(\CS,\alpha(\CS)))}{(i-\alpha(\CS))}$
                    \Else
                        \State $\beta(\CS,i) \gets +\infty$
                    \EndIf
                \EndFor
            \EndFor
            \State $(\hat{\CS},\hat{i}) \gets \arg\min_{\CS,i} \beta(\CS,i)$
        \EndWhile
        \Return $Q$
    \EndFunction
\end{algorithmic}
\end{algorithm}

\section{Performance Analysis}
In order to compare the performance of SDT and PDT approximations, we use numerical simulations. For small networks (e.g. $K \le 5$), it is possible to compare the results with the optimal solution (calculated by the recursive procedure of Algorithm \ref{alg:exh}). However, as $K$ becomes larger, calculating optimal solution becomes computationally intractable; and hence we can only compare SDT and PDT with each other.

For simulations, we choose channel coefficients from a complex Gaussian random variable with zero mean; and normalize the coefficients such that the largest channel amplitude becomes one. The comparison results of the optimal solution (OPT) with respect to PDT and SDT approximations are provided in Table \ref{tab:t4}, for $T_{\lim} =4$ seconds. It can be verified that for small networks, both approximations provide near-optimal performance; such that the difference in QoE sum compared with the optimal solution is less than 0.2\% and 0.6\%, for PDT and SDT respectively. Moreover, the algorithm runtime is improved by at least 93\% and 97\%, for PDT and SDT respectively. Overall, PDT provides an improved result compared with SDT, but also requires a larger runtime.


\begin{table}[H]
    \centering
    \begin{tabular}{|c|c||c|c|c|c|}
        \hline
        \multirow{2}{*}{$K$} & \multirow{2}{*}{$t$} & \multicolumn{2}{c|}{QoE Sum} & \multicolumn{2}{c|}{Algorithm Runtime} \\
        \cline{3-6}
         &  & PDT/OPT & SDT/OPT & PDT/OPT & SDT/OPT \\
        \hline
        \hline
        \multirow{2}{*}{4} & 1 & -0.15\% & -0.51\% & -95.25\% & -98.08\% \\
        \cline{2-6}
         & 2 & -0.04\% & -0.41\% & -93.12\% & -97.18\% \\
        \hline
        \multirow{3}{*}{5} & 1 & -0.08\%  & -0.58\% & -99.86\% & -99.93\% \\
        \cline{2-6}
         & 2 & -0.04\%  & -0.55\% & -99.99\% & -99.99\% \\
        \cline{2-6}
         & 3 & -0.04\% & -0.31\% & -97.37\% & -98.69\% \\
        \hline
    \end{tabular}
    \caption{Performance comparison, $T_{\lim} = 4$ seconds}
    \label{tab:t4}
\end{table}

\begin{figure}[t]
    \centering
    {%

    \begin{tikzpicture}

    \begin{axis}
    [
    width = 0.95\columnwidth,
    height = 0.50\columnwidth,
    axis lines = left,
    xlabel = \smaller {Coded Caching Gain $t$},
    ylabel = \smaller {QoE Improvement [\%]},
    ylabel near ticks,
    xmin=1,
    xmax=10,
    ymax=12,
    ticklabel style={font=\smaller},
    grid=both,
    major grid style={line width=.2pt,draw=gray!30},
    ]
    
    \addplot
    [black]
    table[y=P12,x=t]
    {Data/PDTtoSDT.txt};
    \addlegendentry{\tiny $K = 12$}
    
    \addplot
    [black,dashed]
    table[y=P14,x=t]
    {Data/PDTtoSDT.txt};
    \addlegendentry{\tiny $K = 14$}
    
    \addplot
    [black!70]
    table[y=P16,x=t]
    {Data/PDTtoSDT.txt};
    \addlegendentry{\tiny $K = 16$}
    
    \addplot
    [black!70,dashed]
    table[y=P18,x=t]
    {Data/PDTtoSDT.txt};
    \addlegendentry{\tiny $K = 18$}
    
    \addplot
    [black!50]
    table[y=P20,x=t]
    {Data/PDTtoSDT.txt};
    \addlegendentry{\tiny $K = 20$}
    
    \end{axis}

    \end{tikzpicture}
    }

    \caption{Performance improvement of PDT over SDT}
    \label{fig:PDTtoSDT}
\end{figure}
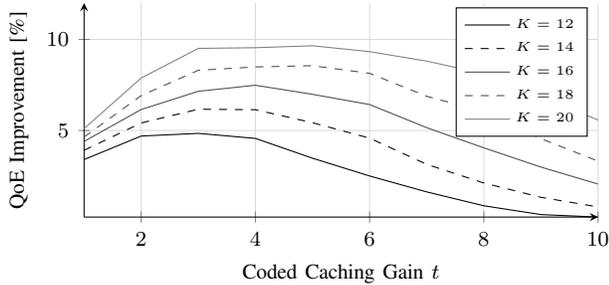

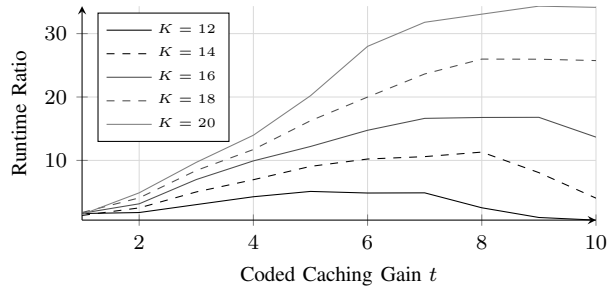
\begin{figure}[t]
    \centering
    {%

    \begin{tikzpicture}

    \begin{axis}
    [
    width = 0.95\columnwidth,
    height = 0.50\columnwidth,
    axis lines = left,
    xlabel = \smaller {Coded Caching Gain $t$},
    ylabel = \smaller {Runtime Ratio},
    ylabel near ticks,
    xmin=1,
    xmax=10,
    legend pos = north west,
    ticklabel style={font=\smaller},
    grid=both,
    major grid style={line width=.2pt,draw=gray!30},
    ]
    
    \addplot
    [black]
    table[y=RT12,x=t]
    {Data/PDTtoSDT.txt};
    \addlegendentry{\tiny $K = 12$}
    
    \addplot
    [black,dashed]
    table[y=RT14,x=t]
    {Data/PDTtoSDT.txt};
    \addlegendentry{\tiny $K = 14$}
    
    \addplot
    [black!70]
    table[y=RT16,x=t]
    {Data/PDTtoSDT.txt};
    \addlegendentry{\tiny $K = 16$}
    
    \addplot
    [black!70,dashed]
    table[y=RT18,x=t]
    {Data/PDTtoSDT.txt};
    \addlegendentry{\tiny $K = 18$}
    
    \addplot
    [black!50]
    table[y=RT20,x=t]
    {Data/PDTtoSDT.txt};
    \addlegendentry{\tiny $K = 20$}
    
    \end{axis}

    \end{tikzpicture}
    }

    \caption{Runtime Increase of PDT over SDT}
    \label{fig:PDTtoSDT_RT}
\end{figure}

In Figures \ref{fig:PDTtoSDT} and \ref{fig:PDTtoSDT_RT} we have compared the performance and runtime of PDT and SDT approximations, for moderate networks with $12 \le K \le 20$ users. In both figures, $T_{\lim}$ is set to 4 seconds. Generally, it can be verified that PDT provides at most 10\% better performance, but requires up to 35 times more computations.
Interestingly, the performance gap is maximum when $\frac{K}{t} \simeq 4$; while the runtime ratio takes its largest value at $\frac{K}{t} \simeq 2$. As the number of $\CS$ sets, i.e. $|\Bar{\CS}| = \binom{K}{t+1}$, is also maximized at $\frac{K}{t} \simeq 2$; this indicates the runtime ratio is proportional to the number of variables in the optimization problem \eqref{eq:optimization_problem_main}. This makes sense, as a larger number of variables necessitates more algorithm iterations; and each iteration in PDT is more complex than SDT. The performance gap does not follow the same rule however. This might be due to the fact that for every set $\CS$, $|\CS| = t+1$; and hence larger $t$ will increase the problem size (and randomness), enabling greedy algorithms like SDT to perform better. This needs more thorough investigation however; which is part of our ongoing research.

Overall, SDT provides a solid performance despite requiring very small computation overhead. Ultimately, selecting the best approximation algorithm depends on the available computation power, as well as network parameters $K$ and $t$.

\section{Conclusion and Future Work}
We proposed a new design approach, based on QoE definition and well-suited to the prominent case of video-based applications, for the worst-user issue of wireless coded caching setups. This approach results in an NP-hard optimization problem, for which we provided two heuristic approximations.

This is a preliminary step for solving the worst-user issue, proposing a new concept which can be further studied in various directions. Using a weighted optimization problem (prioritizing specific users), thorough investigation of approximation algorithms, and extending the concept to multi-antenna setups are few examples of such directions; which are parts of our ongoing research.

\bibliographystyle{IEEEtran}
\bibliography{references}

\end{document}